\documentclass[12pt]{amsart}
\usepackage {amssymb,amsmath}
\makeatletter
\def\@cite#1#2{{\m@th\upshape\bfseries%
[{#1\if@tempswa{\m@th\upshape\mdseries, #2}\fi}]}}
\makeatother
\newtheorem{thm}{Theorem}[section]
\newtheorem{lem}[thm]{Lemma}

\newtheorem{prop}[thm]{Proposition}
\newtheorem{rem}[thm]{Remark}

\newtheorem{defn}[thm]{Definition}
\newtheorem{note}[thm]{Note}

\newtheorem{eg}[thm]{Example}

\newcommand{\Prf}{\noindent\textbf{Proof.\ }}
\newcommand{\bx}{\strut\hfill$\blacksquare$\medbreak}

\newcommand{\ca}{\mathrm{C}^*}

\newenvironment{spmatrix}{\left(\begin{smallmatrix}}{\end{smallmatrix}\right)}

\newcommand{\bbC}{{\mathbb{C}}}

 \newcommand{\A}{{\mathcal{A}}}
 \newcommand{\B}{{\mathcal{B}}}
 \newcommand{\C}{{\mathcal{C}}}
 
 \newcommand{\E}{{\mathcal{E}}}

\renewcommand{\H}{{\mathcal{H}}}

 \newcommand{\K}{{\mathcal{K}}}

 \newcommand{\M}{{\mathcal{M}}}

\renewcommand{\P}{{\mathcal{P}}}
 
 \newcommand{\R}{{\mathcal{R}}}

 \newcommand{\U}{{\mathcal{U}}}

\newcommand{\upchi}{{\raise.35ex\hbox{$\chi$}}}
\newcommand{\fA}{{\mathfrak{A}}}
\newcommand{\fB}{{\mathfrak{B}}}

\newcommand{\qand}{\quad\text{and}\quad}

\newcommand{\qfor}{\quad\text{for}\quad}

\newcommand{\qforal}{\quad\text{for all}\quad}

\newcommand{\Alg}{\operatorname{Alg}}

\newcommand{\spn}{\operatorname{span}}

\newcommand{\Tr}{\operatorname{Tr}}

\newcommand{\fix}{\operatorname{Fix}}
\def\bra#1{\langle #1|}
\def\ket#1{|#1 \rangle}
\def\kb#1#2{|#1\rangle\!\langle #2 |}
\def\one{{\mathchoice{\rm 1\mskip-4mu l}{\rm 1\mskip-4mu l}{\rm 1\mskip-4.5mu l}{\rm
1\mskip-5mu l}}}

\begin{document}

%%%%%%%%%%%%%%%%%%%%%%%%%%%%%%%%%%%%%%%%%%%%%%
%%%%%%%%%%
\title[Operator Quantum Error Correction]{Operator Quantum Error
Correction}
%\thanks{draft July, 2003}
%
\author[D.W. Kribs, R. Laflamme, D. Poulin, M. Lesosky]{David W. Kribs$^{1,2}$, Raymond Laflamme$^{2,3}$,  David Poulin$^{2,4}$, and Maia Lesosky$^1$}
%\thanks{2000 {\it Mathematics Subject Classification.} 47L90,  81P68.}
%\thanks{{\it key words and phrases.} quantum operation, generalized noiseless subsystem,
%decoherence-free subspace, operator quantum error correction.}

\address{$^1$Department of Mathematics and Statistics,
University of Guelph, Guelph, Ontario, Canada  N1G 2W1}
\address{$^2$Institute for Quantum Computing, University of
Waterloo, Waterloo, ON, CANADA N2L 3G1} \address{$^3$Perimeter
Institute for Theoretical Physics, 31 Caroline St. North,
Waterloo, ON, CANADA N2L 2Y5}
\address{$^4$School of Physical Sciences, The University of Queensland, QLD 4072, Australia}

%\date{}
%
\begin{abstract}
This paper is an expanded and more detailed version of the work
\cite{KLP04} in which the Operator Quantum Error Correction
formalism was introduced. This is a new scheme for the error
correction of quantum operations that incorporates the known
techniques --- i.e. the standard error correction model, the
method of decoherence-free subspaces, and the noiseless subsystem
method --- as special cases, and relies on a generalized
mathematical framework for noiseless subsystems that applies to
arbitrary quantum operations. We also discuss a number of examples
and introduce the notion of ``unitarily noiseless subsystems''.
\end{abstract}
\maketitle
%%%%%%%%%%%%%%%%%%%%%%%%%%%%%%%%%%%%%%%%%%%%%%
%%%%%%%%%%%

%%%%%%%%%%%%%%%%%%%%%%%%%%%%%%%%%%%%%%
%\section{Introduction}\label{S:intro}
%%%%%%%%%%%%%%%%%%%%%%%%%%%%%%%%%%%%%%

A unified and generalized approach to quantum error correction,
called Operator Quantum Error Correction (OQEC), was recently
introduced in \cite{KLP04}. This formalism unifies all of the
known techniques for the error correction of quantum operations --
i.e. the standard model \cite{Sho95a,Ste96a,BDSW96a,KL97a}, the
method of decoherence-free subspaces
\cite{PSE96,DG97c,ZR97c,LCW98a} and the noiseless subsystem method
\cite{KLV00a,Zan01b,KBLW01a} -- under a single umbrella. An
important new framework introduced as part of this scheme opens up
the possibility of studying noiseless subsystems for arbitrary
quantum operations.

This paper is an expanded and more detailed version of the work
\cite{KLP04}. We provide complete details for proofs sketched
there, and in some cases we present an alternative ``operator''
approach that leads to new information. Specifically, we show that
correction of the general codes introduced in \cite{KLP04} is
equivalent to correction of certain operator algebras, and we use
this  to give a new proof for the main testable conditions in this
scheme. In addition, we discuss a number of examples throughout
the paper, and introduce the notion of ``unitarily noiseless
subsystems'' as a relaxation of the requirement in the noiseless
subsystem formalism for immunity to errors.

%We also connect this work with aspects of more recent OQEC related
%efforts. In particular, we show that the fundamental formula in
%the formulation of the ``Quantum Computer Condition'' recently
%introduced in \cite{GHT05} is captured as a special case of the
%UNS framework.

%%%%%%%%%%%%%%%%%%%%%%%%%%%%%%%%%%%%%%
\section{Preliminaries}\label{S:prelim}
%%%%%%%%%%%%%%%%%%%%%%%%%%%%%%%%%%%%%%

%%%%%%%%%%%%%%%%%%%%%%%%%%%%%%%%%%%%%%
\subsection{Quantum Operations}\label{S:channels}
%%%%%%%%%%%%%%%%%%%%%%%%%%%%%%%%%%%%%%

Let $\H$ be a (finite-dimensional) Hilbert space and let $\B(\H)$
be the set of operators on $\H$. A {\it quantum operation} (or
{\it channel}, or {\it evolution}) on $\H$ is a linear map
$\E:\B(\H)\rightarrow\B(\H)$ that is completely positive and
preserves traces. Every channel has an ``operator-sum
representation'' of the form $ \E(\sigma) = \sum_a E_a \sigma
E_a^\dagger$, $\forall \sigma \in \B(\H)$, where $\{E_a\}\subseteq
\B(\H)$ are the Kraus operators (or errors) associated with $\E$.
As a convenience we shall write $\E = \{E_a \}$ when the $E_a$
determine $\E$ in this way.

The choice of operators that yield this form is not unique, but if
$\E=\{E_a\} =\{F_b\}$ (without loss of generality assume the
cardinalities of the sets are the same), then there is a unitary
matrix $U = (u_{ab})$ such that $ E_a = \sum_b u_{ab} F_b$
$\forall \,a$. The map $\E$ is said to be {\it unital} or {\it
bistochastic} if $\E(\one) = \sum_a E_a E_a^\dagger = \one$. Trace
preservation of $\E$ can be phrased in terms of the error
operators via the equation $ \sum_a E_a^\dagger E_a = \one$, which
is  equivalent to the dual map for $\E$ being unital.

%%%%%%%%%%%%%%%%%%%%%%%%%%%%%%%%%%%%%%
\subsection{Standard Model for Quantum Error Correction}\label{S:standardmodel}
%%%%%%%%%%%%%%%%%%%%%%%%%%%%%%%%%%%%%%

The ``Standard Model'' for the error correction of quantum
operations \cite{Sho95a,Ste96a,BDSW96a,KL97a} consists of  triples
$(\R,\E,\C)$ where $\C$ is a subspace, a {\it quantum code}, of a
Hilbert space $\H$ associated with a given quantum system. The
error $\E$ and recovery $\R$ are quantum operations on $\B(\H)$
such that $\R$ undoes the effects of $\E$ on $\C$ in the following
sense:
\begin{eqnarray}\label{reverse}
(\R\circ \E) \, (\sigma) = \sigma \quad \forall\, \sigma = P_\C
\sigma P_\C,
\end{eqnarray}
where $P_\C$ is the projection of $\H$ onto $\C$.

When there exists such  an $\R$ for a given pair $\E,\C$, the
subspace $\C$ is said to be {\it correctable for $\E$}. The
existence of a recovery operation $\R$ of $\E=\{E_a\}$ on $\C$ may
be cleanly phrased in terms of the $\{E_a\}$ as follows
\cite{BDSW96a,KL97a}:
\begin{equation}
P_\C E_a^\dagger E_b P_\C = \lambda_{ab}P_\C \quad \forall\, a,b
\label{eq:standard}
\end{equation}
for some matrix $\Lambda = (\lambda_{ab})$. It is easy to see that
this condition is independent of  the operator-sum representation
for $\E$.

%%%%%%%%%%%%%%%%%%%%%%%%%%%%%%%%%%%%%%
\subsection{Noiseless Subsystems and Decoherence-Free
Subspaces}\label{S:noiseless}
%%%%%%%%%%%%%%%%%%%%%%%%%%%%%%%%%%%%%%

Let $\E = \{ E_a\}$ be a quantum operation on $\H$. Let $\A$ be
the $\ca$-algebra generated by the $E_a$, so $ \A = \Alg
\{E_a,E_a^\dagger\}. $ This is the set of  polynomials in the
$E_a$ and $E_a^\dagger$. As a $\dagger$-algebra  (i.e., a
finite-dimensional $\ca$-algebra \cite{Arv76a,Dav96a,Tak79a}),
$\A$ has a unique decomposition up to unitary equivalence of the
form
\begin{eqnarray}\label{decomp}
\A &\cong& \bigoplus_J \big(\M_{m_J} \otimes \one_{n_J} \big),
\end{eqnarray}
where $\M_{m_J}$ is the full matrix algebra $\B(\bbC^{m_J})$
represented with respect to a given orthonormal basis and
$\one_{n_J}$ is the identity on $\bbC^{n_J}$. This means there is
an orthonormal basis such that the matrix representations of
operators in $\A$ with respect to this basis have the form in
Eq.~(\ref{decomp}). Typically $\A$ is called the {\it interaction
algebra} associated with the operation $\E$.

The standard ``noiseless subsystem'' method of quantum error
correction \cite{KLV00a,Zan01b,KBLW01a} makes use of the operator
algebra structure of the {\it noise commutant} associated with
$\E$;
\[
\A' = \big\{\sigma\in\B(\H): E \sigma = \sigma E \,\,\,\forall
E\in\{E_a,E_a^\dagger\} \big\}.
\]
Observe that when $\E$ is unital, all the states encoded in
$\A^\prime$ are immune to the errors of $\E$. Thus, this is in
effect a method of passive error correction. The structure of $\A$
given in Eq.~(\ref{decomp}) implies that the noise commutant is
unitarily equivalent to
\begin{eqnarray}\label{commdecomp}
\A'&\cong& \bigoplus_J \big(\one_{m_J} \otimes \M_{n_J}\big).
\end{eqnarray}

It is obvious from Eqs.~(\ref{decomp},\ref{commdecomp}) that
elements of $\A^\prime$ are immune to the errors of $\A$ when $\E$
is unital. In \cite{Kri03a} the converse of this statement was
proved. Specifically, when $\E$ is unital the noise commutant
coincides with the fixed point set for $\E$; i.e.,
\begin{eqnarray}\label{fixedptthm}
\A' = \fix(\E) &=& \{ \sigma\in\B(\H): \E(\sigma)=\sum_a E_a\sigma
E_a^\dagger = \sigma\}.
\end{eqnarray}
This is precisely the reason that $\A^\prime$ may be used to
produce noiseless subsystems for unital $\E$. We note that the
noiseless subsystem  method may be regarded as containing the
method of {\it decoherence-free
subspaces}~\cite{PSE96,DG97c,ZR97c,LCW98a} as a special case, in
the sense that this method makes use of the summands $
\one_{m_J}\otimes\M_{n_J}$ where $m_J=1$, inside the noise
commutant $\A^\prime$ for encoding information.

While many physical noise models satisfy the unital constraint,
the generic quantum operation is non-unital. Below we show how
shifting the focus from $\A^\prime$ to $\fix(\E)$ (and related
sets) quite naturally leads to the notion of noiseless subsystems
that applies to arbitrary quantum operations.

%%%%%%%%%%%%%%%%%%%%%%%%%%%%%%%%%%%%%%%%%%%%%%%%%%%%%%%%%%%%%
\section{Noiseless Subsystems For Arbitrary Quantum Operations}\label{S:genNS}
%%%%%%%%%%%%%%%%%%%%%%%%%%%%%%%%%%%%%%%%%%%%%%%%%%%%%%%%%%%%%

In this section we describe a generalized mathematical framework
for noiseless subsystems that applies to arbitrary (not
necessarily unital) quantum operations and serves as a building
block for the OQEC scheme presented below. Note that a subsystem
that is noiseless for a certain map will also be noiseless for any
other map whose Kraus operators are linear combinations of the
Kraus operators of the original map. Hence, for the purpose of
noiseless encoding, any map whose Kraus operators span is closed
under conjugation is equivalent to a unital map. The mathematical
framework utilized in \cite{KLV00a,Zan01b,KBLW01a} produces
noiseless subsystems for precisely these kinds of operations, and
so may effectively be regarded as restricted to unital channels.
That being said, it is desirable to find a means by which
noiseless subsystems can be discovered without relying on the
unital nature of an operation, or the structure of its noise
commutant. The main result of this section (Theorem~\ref{thm:NS})
shows explicitly how this may be accomplished.
%However, the
%standard mathematical framework either explicitly focuses on
%unital maps, or does so implicitly by relying on the algebraic
%approach outlined above. The results presented in \cite{KBLW01a}
%apply to an arbitrary map in the Markovian approximation, and also
%yield an algebraic structure.

Note that the structure of the algebra
$\A$ given in Eq.~(\ref{decomp}) induces a natural decomposition
of the Hilbert space
\begin{equation*}
\H = \bigoplus_J \H^A_J\otimes\H^B_J,
\end{equation*}
where the ``noisy subsystems" $\H^A_J$ have dimension $m_J$ and
the ``noiseless subsystems" $\H^B_J$ have dimension $n_J$. For
brevity, we focus on the case where information is encoded in a
single noiseless sector of $\B(\H)$, and hence
\begin{equation*}
\H = (\H^A \otimes \H^B) \oplus \K \label{eq:decomp}
\end{equation*}
with $\dim(\H^A) = m$, $\dim(\H^B) = n$ and $\dim \K=\dim\H - mn$.
We shall write $\sigma^A$ for operators in $\B(\H^A)$ and
$\sigma^B$ for operators in $\B(\H^B)$. Thus the restriction of
the noise commutant $\A'$ to $\H^A\otimes\H^B$ consists of the
operators of the form $\sigma = \one^A\otimes\sigma^B$ where
$\one^A$ is the identity element of $\B(\H^A)$.

For notational purposes, assume that ordered orthonormal bases
have been chosen for $\H^A =
{\mathrm{span}}\{\ket{\alpha_i}\}_{i=1}^m$ and $\H^B =
{\mathrm{span}}\{\ket{\beta_k}\}_{k=1}^n$ that yield the matrix
representation of the corresponding subalgebra of $\A'$ as $\one^A
\otimes \B(\H^B)\cong\one_m \otimes \M_n$. We let
\begin{equation}\label{matrixunitdefn}
 P_{kl} \equiv \kb{\alpha_k}{\alpha_l}\otimes\one^B \quad \forall\, 1\leq k,l \leq
m
\end{equation}
denote the corresponding family of ``matrix units'' in $\A$
associated with this decomposition. The following identities are
readily verified and are the defining properties for a family of
matrix units:
\begin{eqnarray*}
P_{kl} & =& P_{kk} P_{kl} P_{ll}  \quad \forall\, 1\leq k,l\leq m \\
P_{kl}^\dagger &=& P_{lk} \quad\quad\quad \forall\, 1\leq k,l \leq m \\
P_{kl}P_{l'k'} &=& \left\{ \begin{array}{cl} P_{kk'} & \mbox{if
$l=l'$} \\ 0 & \mbox{if $l\neq l'$} \end{array}\right..
\end{eqnarray*}
Define the projection $P_\fA \equiv P_{11} +\ldots + P_{mm}$, so
that $P_\fA\H = \H^A\otimes\H^B$, $P_\fA^\perp = \one - P_\fA$ and
$P_\fA^\perp\H = \K$.  Further define a superoperator $\P_\fA$ by
the action $\P_\fA(\cdot) = P_\fA(\cdot) P_\fA$. The following
result is readily proved.

\begin{lem}
The map $\Gamma : \B(\H)\rightarrow\B(\H)$ given by $\Gamma =
\{P_{kl}\}$ satisfies the following:
\begin{equation}
\Gamma(\sigma) = \sum_{k,l}P_{kl}\sigma P_{kl}^\dagger = \one^A
\otimes (\Tr_A\circ\P_\fA)(\sigma) \in \one^A\otimes\B(\H^B),
\end{equation}
for all operators $\sigma \in \B(\H)$, so in particular
$\Gamma(\sigma^A\otimes\sigma^B) \propto \one^A\otimes \sigma^B$
for all $\sigma^A$ and $\sigma^B$. \label{lemma:map}
\end{lem}

\begin{note}
{\rm While we have stated this result as part of a discussion on a
subalgebra of a noise commutant, it is valid for any
$\dagger$-algebra $\fB\cong \one^A \otimes \B(\H^B)$ with matrix
units $\{P_{kl}\}$ generating the algebra
$\B(\H^A)\otimes\one^B$.}
\end{note}

We now turn to the generalized noiseless subsystems method. In
this framework, the quantum information is encoded in $\sigma^B$;
i.e., the state of the noiseless subsystem. But it is not
necessary for the noisy subsystem to remain in the maximally mixed
state $\one^A$ under $\E$, as is the case for noiseless subsystems
of unital channels, it could in principle get mapped to any other
state.

In order to formalize this idea, define for a fixed decomposition
$\H = (\H^A\otimes\H^B) \oplus\K$ the set of operators
\begin{eqnarray}\label{eq:semigroup}
\fA = \{\sigma\in\B(\H) : \sigma = \sigma^A\otimes\sigma^B,\,{\rm
for\,\,some}\,\, \sigma^A\,\,{\rm and}\,\, \sigma^B\}.
\end{eqnarray}
Notice that this set has the structure of a semigroup and includes
operator algebras such as $\fA_0 \equiv \one^A\otimes\B(\H^B)$ and
$\kb{\alpha_k}{\alpha_k}\otimes\B(\H^B)$. We note that in the
formulation below, the operation $\E$ maps the set of operators on
the subspace $P_\fA\H = \H^A\otimes\H^B$ to itself.

\begin{lem}\label{NSlemma}
Given a fixed decomposition $\H = (\H^A\otimes\H^B) \oplus\K$ and
a quantum operation $\E$ on $\B(\H)$, the following four
conditions are equivalent, and are the defining properties of the
noiseless subsystem $B$:
\begin{enumerate}
\item[({\it 1})] $\forall\sigma^A\ \forall\sigma^B,\ \exists
\tau^A\ :\ \E(\sigma^A\otimes\sigma^B) = \tau^A\otimes\sigma^B$
\item[({\it 2})] $ \forall\sigma^B,\ \exists \tau^A\ :\
\E(\one^A\otimes\sigma^B) = \tau^A \otimes \sigma^B$ \item[({\it
3})] $\forall\sigma\in \fA\ :\ \big(\Tr_A\circ \P_\fA\circ
\E\big)(\sigma) =\Tr_A(\sigma)$.
\end{enumerate}
\label{lemma:generalNS}
\end{lem}

\Prf The implications {\it 1.} $\Rightarrow$ {\it 2.} and {\it 1.}
$\Rightarrow$ {\it 3.} are trivial. To prove {\it 2.}
$\Rightarrow$ {\it 1.}, first let $\ket{\psi}\in\H^B$ and put $P =
\kb{\psi}{\psi}$. Suppose that $\{\ket{\alpha_k}\}$ is an
orthonormal basis for $\H^A$. Then $\sum_{k=1}^m
\kb{\alpha_k}{\alpha_k} = \one^A$ and by {\it 2.} and the
positivity of $\E$ we have for all $k$,
\begin{eqnarray*}
0 \leq \E( \kb{\alpha_k}{\alpha_k} \otimes P)  &\leq& \E (\one^A
\otimes P) \\ &=& \tau^A \otimes P \\   &=& (\one^A\otimes P)
(\tau^A \otimes P)(\one^A \otimes P).
\end{eqnarray*}
It follows that there are positive operators $\sigma_{\psi,k}\in
\B(\H^A)$ such that $\E(\kb{\alpha_k}{\alpha_k}\otimes P) =
\sigma_{\psi,k} \otimes P$ for all $k$. A standard linearity
argument may be used to show that the operators $\sigma_{\psi,k}$
do not depend on $\ket{\psi}$. Condition~{\it 1.} now follows from
the linearity of $\E$.

%To verify this claim,  for clarity we shall suppose that $\dim
%\H^B =2$. The case of general $\H^B$ easily follows. So let
%$\ket{\psi_i}$, $i=1,2$, be an orthonormal basis for $\H^B$. Let
%$P_i = \kb{\psi_i}{\psi_i}$, $i=1,2$, and put $P_{\pm} =
%\kb{\pm}{\pm}$ where $\ket{\pm} = \frac{1}{\sqrt{2}}(\ket{\psi_1}
%\pm \ket{\psi_2})$. Fix $\alpha=\alpha_k$. By the above argument,
%there are operators $\sigma_{\pm,\alpha}$ and $\sigma_{i,\alpha}$
%on $\H^A$ such that
%\[
%\E(\kb{\alpha}{\alpha}\otimes P_{\pm}) = \sigma_{\pm,\alpha}
%\otimes P_{\pm} \qand \E(\kb{\alpha}{\alpha}\otimes P_{i}) =
%\sigma_{i,\alpha} \otimes P_{i}.
%\]
%In particular, as $\one^B = P_+ + P_- = P_1 + P_2$, we have
%\begin{eqnarray*}
% \E(\kb{\alpha}{\alpha} \otimes \one^B) &=&
%\sigma_{1,\alpha} \otimes P_{1} + \sigma_{2,\alpha} \otimes P_{2}
%\\ &=& \sigma_{+,\alpha} \otimes P_{+} + \sigma_{-,\alpha} \otimes
%P_{-}.
%\end{eqnarray*}
%If we compress this equation by the projection $\one^A \otimes
%P_1$, we obtain
%\begin{eqnarray*}
%(\one^A\otimes P_1) \E(\kb{\alpha}{\alpha} \otimes \one^B) (\one^A
%\otimes P_1) &=& \sigma_{1,\alpha} \otimes P_{1}
%\\ &=& \frac{1}{2} (\sigma_{+,\alpha} + \sigma_{-,\alpha}) \otimes
%P_{1}.
%\end{eqnarray*}
%Thus, $\sigma_{1,\alpha} =  \frac{1}{2} (\sigma_{+,\alpha} +
%\sigma_{-,\alpha})$ and since the same identity holds for
%$\sigma_{2,\alpha}$ when we compress by $\one^A\otimes P_2$, we
%obtain $\sigma_{1,\alpha} = \sigma_{2,\alpha}$. As $\ket{\alpha}$
%and $\ket{\psi_i}$, $i=1,2$, were chosen arbitrarily, the claim
%holds.

To prove {\it 3.} $\Rightarrow$ {\it 2.}, first note that since
$\E$ and $\Tr_A$ are positive and trace preserving, {\it 3}.\@
implies that $\big(\P_\fA \circ \E\big)(\sigma) = \E(\sigma)$ for
all $\sigma \in \fA$. Now fix $\ket{\psi}\in\H^B$ and put $\sigma
= \one^A\otimes P$ where $P = \kb{\psi}{\psi}$. Then by {\it 3.}
we have
\[
\Tr_A \big((\one^A\otimes P) \,\E(\sigma)\,(\one^A\otimes P)\big)
= \Tr_A (\sigma).
\]
It follows  again from the trace preservation and positivity of
$\Tr_A$ and $\E$ that $\sigma \E(\sigma) \sigma = \E(\sigma)$, and
hence there is a $\tau^A$ such that $\E(\sigma) = \tau^A \otimes
P$. The above argument may now be used to show that $\tau^A$ is
independent of $\ket{\psi}$, and the rest follows from the
linearity of $\E$. \bx

\begin{defn}\label{defn:NS}
{\rm The subsystem $B$ is said to be {\it noiseless} for $\E$ when
it satisfies one --- and hence all --- of the conditions in
Lemma~\ref{lemma:generalNS}.}
\end{defn}

%Note that the generalized definition of noiseless subsystems
%coincides with the standard definition when $\dim(\H^A) = 1$.
%Hence, the notion of decoherence-free subspaces is not altered by
%this generalization.

We next give necessary and sufficient conditions for a subsystem
to be noiseless for a map $\E=\{E_a\}$.

\begin{thm}\label{thm:NS}
Let $\E = \{E_a\}$ be a quantum operation on $\B(\H)$ and let
$\fA$ be a semigroup in $\B(\H)$ as in Eq.~(\ref{eq:semigroup}).
Then the following three conditions are equivalent:
\begin{itemize}
\item[({\it 1})] The $B$-sector of $\fA$ encodes a noiseless
subsystem for $\E$ (decoherence-free subspace in the case m=1), as
in Definition~\ref{defn:NS}. \item[({\it 2})] The subspace $P_\fA
\H = \H^A \otimes \H^B$ is invariant for the operators $E_a$ and
the restrictions $E_a|_{P_\fA\H}$ belong to the algebra
$\B(\H^A)\otimes\one^B$. \item[({\it 3})] The following two
conditions hold for any choice of matrix units $\{P_{kl}: 1 \leq
k,l \leq m \}$ for $\B(\H^A)\otimes\one^B$ as in
Eq.~(\ref{matrixunitdefn}):
\begin{equation}
P_{kk} E_a P_{ll} = \lambda_{akl} P_{kl} \quad\forall\, a,k,l
\label{eq:cond1}
\end{equation}
for some set of scalars $(\lambda_{akl})$ and
\begin{equation}
 E_a P_\fA = P_\fA E_a P_\fA \quad\forall\, a. \label{eq:cond2}
\end{equation}
\end{itemize}
\end{thm}

\Prf Since the matrix units $\{P_{kl}\}$ generate $\B(\H^A)\otimes
\one^B$ as an algebra, it follows that {\it 3.} is a restatement
of {\it 2.} To prove the necessity of
Eqs.~(\ref{eq:cond1},\ref{eq:cond2}) for {\it 1.}, let $\Gamma:
\B(\H)\rightarrow \one^A \otimes \B(\H^B)$ be defined by the
matrix units for $\fA$ as above and note that
Lemma~\ref{lemma:map} and Lemma~\ref{lemma:generalNS} imply
\begin{equation}
\big(\Gamma\circ\E\circ\Gamma\big)(\sigma) \propto \Gamma(\sigma)
\qforal \sigma \in \B(\H). \label{eq:conditionNS}
\end{equation}
As in the proof of Lemma~\ref{lemma:generalNS}, the
proportionality factor cannot depend on $\sigma$, so the sets of
operators $\{P_{ki}E_aP_{jl}\}$ and $\{\lambda P_{k'l'}\}$ define
the same map for some scalar $\lambda$. We may thus find a set of
scalars $\mu_{kiajl,k'l'}$ such that
\begin{equation}
P_{ki}E_aP_{jl} = \sum_{k'l'} \mu_{kiajl,k'l'} P_{k'l'}.
\end{equation}
Multiplying both sides of this equality on the  right by $P_l$ and
on the left by $P_k$, we see that $\mu_{kiajl,k'l'} = 0$ when
$k\neq k'$ or $l\neq l'$. This implies Eq.~(\ref{eq:cond1}) with
$\lambda_{akl} = \mu_{kkall,kl}$.

For the second condition, note that as a consequence of
Lemma~\ref{lemma:generalNS}, we have $P_\fA^\perp \E(P_\fA
(\sigma)) P_\fA^\perp = 0$ for all $\sigma \in \B(\H)$.
Equation~(\ref{eq:cond2}) follows from this observation via
consideration of the operator-sum representation (see
$\S$~\ref{S:channels}) for $\E$.

To  prove sufficiency of Eqs.~(\ref{eq:cond1}), (\ref{eq:cond2})
for {\it 1.}, we use the identity $P_\fA = \sum_{k=1}^m P_k$ to
establish for all $\sigma= P_\fA \sigma \in\fA$,
\begin{eqnarray*}
\E(\sigma) &=& (P_\fA + P_\fA^\perp) \sum_a E_a\sigma E^\dagger_a(P_\fA + P_\fA^\perp) \\
&=&\sum_a P_\fA E_a \sigma E^\dagger_a P_\fA \\
&=& \sum_{a,k,k'} P_{kk}E_a\sigma E_a^\dagger P_{k'k'}.
\end{eqnarray*}
Combining this with the identity
\[
\sigma^A\otimes\sigma^B= P_\fA( \sigma^A\otimes\sigma^B) P_\fA =
\sum_{l,l'} P_{ll} (\sigma^A \otimes \sigma^B) P_{l'l'}
\]
implies for all $\sigma = \sigma^A\otimes\sigma^B\in\fA$,
\begin{eqnarray*}
\E(\sigma^A\otimes\sigma^B) &=& \sum_{a,k,k',l,l'} P_{kk}E_aP_{ll}(\sigma^A\otimes\sigma^B) P_{l'l'} E^\dagger_a P_{k'k'} \\
&=& \sum_{a,k,k',l,l'} \lambda_{akl}\overline{\lambda}_{ak'l'}
P_{kl} (\sigma^A\otimes\sigma^B) P_{l'k'}.
\end{eqnarray*}
The proof now follows from the fact that the matrix units $P_{kl}$
act trivially on the $\B(\H^B)$ sector. \bx

\begin{rem}
{\rm In the case that the semigroup $\fA$ is determined by a
matrix block inside the noise commutant $\A^\prime$ for a unital
channel $\E = \{E_a\}$, and hence arises through the algebraic
approach as in the discussion at the start of this section, the
conditions Eqs.~(\ref{eq:cond1},\ref{eq:cond2}) follow from the
structure of $\A  = \Alg \{ E_a, E_a^\dagger \}$ determined by the
matrix units $P_{kl}$. However,
Eqs.~(\ref{eq:cond1},\ref{eq:cond2}) do not necessarily imply that
the noiseless subsystem $B$ is obtained via the noise commutant
for $\E$. See \cite{CK05} for further discussions on this point.
%
%operators of $\fA$ are in the commutant of the interaction algebra
%$\A$. In fact, recent work \cite{CK05} gives a method to find all
%noiseless subsystems  for arbitrary quantum operations. It is
%shown that the noise commutant still yields noiseless subsystems
%(even in the non-unital case), and that all other noiseless
%subsystems are shown to arise through an interplay between the
%noise commutant and projections $P$ that satisfy the equation
%$\E(P) = P\E(P)P$.
}
\end{rem}

We now discuss a pair of non-unital examples of channels with
noiseless subsystems.

\begin{eg}
{\rm As a simple illustration of a noiseless subsystem in a
non-unital case, consider the quantum channel $\E : \M_4
\rightarrow \M_4$ with errors $\E = \{ E_1,E_2\}$ obtained as
follows. Fix $\gamma$, $0 \leq \gamma \leq 1$, and with respect to
the basis $\{\ket{0},\ket{1}\}$ let
\[
F_0 = \left(\begin{matrix} \sqrt\gamma & 0 \\ 0 & \sqrt{1-\gamma}
\end{matrix}\right) \qand
F_1 = \left(\begin{matrix} 0 & \sqrt{\gamma} \\ \sqrt{1-\gamma} & 0
\end{matrix}\right).
\]
Then define $E_i = F_i \otimes\one_2$, for $i=0,1$. That $\sum_i E_i^\dagger E_i = \one_4$ follows from $\sum_i F_i^\dagger F_i = \one_2$, which can be verified straightforwardly.

Decompose $\bbC^4 = \H^A \otimes \H^B$ with respect to the
standard basis, so that $\H^A = \H^B = \bbC^2$.  Then for all
$\sigma = \sigma^A \otimes \sigma^B$, we have
\[
\E(\sigma) = \sum_{i=0}^1 E_i(\sigma^A \otimes \sigma^B)
E_i^\dagger = \Big( \sum_{i=0}^1 F_i \sigma^A F_i^\dagger \Big)
\otimes \sigma^B.
\]
The operator $\tau^A$ from Lemma~\ref{lemma:generalNS} is given by
$\tau^A = \sum_i F_i \sigma^A F_i^\dagger$ in this case. It
follows that $B$ encodes a noiseless subsystem for $\E$.  Also
observe that, as opposed to the completely error-free evolution
that characterizes the unital case, in this case we have
$\E(\one^A\otimes\sigma^B) \neq \one^A\otimes\sigma^B$.}
\end{eg}

\begin{eg}
{\rm We next present a non-unital channel with a pair of noiseless
subsystems; one that is supported by the noise commutant, and one
that is not. We shall explicitly indicate
Eqs.~(\ref{eq:cond1},\ref{eq:cond2}) in this case. Let $\E = \{
E_0 , E_1\}$ be the channel on $\bbC^4 = \bbC^2\otimes\bbC^2$ with
Kraus operators defined with respect to the computational basis by
\begin{eqnarray*}
E_0 &=& \alpha \big( \kb{00}{00} + \kb{11}{11}\big) + \kb{01}{01}
+ \kb{10}{10}, \\ E_1 &=& \beta \big( \kb{00}{00} + \kb{10}{00} +
\kb{01}{11} + \kb{11}{11} \big),
\end{eqnarray*}
where $0<q<1$ is fixed, and $\alpha = \sqrt{1-2q}$ and
$\beta=\sqrt{q}$. (Notice that $\E$ is non-unital; $\E(\one)\neq
\one$.)

Let $\H^{B_1} = \spn\{ \ket{01},\ket{10}\}$ and $\H^{A_1} = \bbC$,
so that $\H^{A_1}\otimes\H^{B_1} = \H^{B_1}$. We may regard
$\ket{0_L} = \ket{01}$ and $\ket{1_L} = \ket{10}$ as logical zero
and logical one states in this case. Let $Q = \kb{01}{01} +
\kb{10}{10}$. Then
\[
E_0 Q = Q = Q E_0 = Q E_0 Q
\]
\[
E_1 Q = 0 = Q E_1 Q.
\]
Thus, Eqs.~(\ref{eq:cond1},\ref{eq:cond2}) are satisfied and it
follows from Theorem~\ref{thm:NS} that $B_1$ is a noiseless
subsystem (a subspace in this case) for $\E$. To see this
explicitly, let $\sigma\in\B(\H^{B_1})$ be arbitrary, and so
\[
\sigma = a \kb{01}{01} +  b \kb{01}{10} + c \kb{10}{01} + d
\kb{10}{10},
\]
for some $a,b,c,d\in\bbC$. Then
\[
\E(\sigma) = E_0\sigma E_0^\dagger + E_1 \sigma E_1^\dagger =
\sigma,
\]
and the conditions of Lemma~\ref{NSlemma} are satisfied for all
$\sigma\in\B(\H^{B_1}) = \B(\H^{A_1}\otimes\H^{B_1})$. Observe
that a typical operator $\sigma\in\B(\H^{B_1})$ satisfies
$E_1\sigma = 0 \neq \sigma E_1$, and hence this noiseless
subsystem is not supported by the noise commutant for $\E$.

There is another noiseless subsystem for $\E$ which is supported
by the noise commutant. Decompose $\bbC^4=\H^{A_2}\otimes\H^{B_2}$
into the product of a pair of single qubit systems $\H^{A_2} =
\spn\{\ket{\alpha_1},\ket{\alpha_2}\} = \bbC^2$ and $\H^{B_2} =
\spn\{\ket{\beta_1},\ket{\beta_2}\} = \bbC^2$ such that
\begin{eqnarray*}
\ket{\alpha_1}\otimes\ket{\beta_1} &=&
\frac{\ket{00}+\ket{11}}{\sqrt{2}} \\
\ket{\alpha_1}\otimes\ket{\beta_2} &=&
\frac{\ket{00}-\ket{11}}{\sqrt{2}}
\\\ket{\alpha_2}\otimes\ket{\beta_1} &=&
\frac{\ket{10}+\ket{01}}{\sqrt{2}} \\
\ket{\alpha_2}\otimes\ket{\beta_2} &=&
\frac{\ket{10}-\ket{01}}{\sqrt{2}} .
\end{eqnarray*}
As noted below, $\ket{0_L}=\ket{\beta_1}$ and
$\ket{1_L}=\ket{\beta_2}$ are logical zero and logical one states
that remain immune to the errors of $\E$. For $1\leq k,l \leq 2$,
let
\begin{eqnarray*}
P_{kl} &=& \kb{\alpha_k}{\alpha_l} \otimes \one^{B_2} \\
&=& \kb{\alpha_k}{\alpha_l} \otimes (  \kb{\beta_1}{\beta_1}+ \kb{\beta_2}{\beta_2}    ) \\
&=&
(\ket{\alpha_k}\otimes\ket{\beta_1})(\bra{\alpha_l}\otimes\bra{\beta_1})
+
(\ket{\alpha_k}\otimes\ket{\beta_2})(\bra{\alpha_l}\otimes\bra{\beta_2})
\end{eqnarray*}
be the matrix units associated with this decomposition, and notice
that these operators are given by
\[
P_{11} = \kb{00}{00} + \kb{11}{11} \quad P_{12} = \kb{00}{10} +
\kb{11}{01}
\]
\[
P_{21} = \kb{10}{00} + \kb{01}{11} \quad P_{22} = \kb{10}{10} +
\kb{01}{01}.
\]

We calculate to find:
\[
P_{11} E_0 P_{11} = \alpha P_{11} \quad P_{11} E_0 P_{22} = 0
P_{12}
\]
\[
P_{22} E_0 P_{11} = 0 P_{21} \quad P_{22} E_0 P_{22} = P_{22}
\]
\[
P_{11} E_1 P_{11} = \beta P_{11} \quad P_{11} E_1 P_{22} = 0
P_{12}
\]
\[
P_{22} E_1 P_{11}  = 0 P_{21} \quad P_{22} E_1 P_{22} = 0 P_{22}.
\]
Thus, Eqs.~(\ref{eq:cond1},\ref{eq:cond2}) are satisfied and it
follows from Theorem~\ref{thm:NS} that $B_2$ is a noiseless
subsystem for $\E$. As an illustration of the conditions from
Lemma~\ref{NSlemma} in this case, one can check that
\[
\E\big(\one_2 \otimes \sigma \big) = \left( \begin{smallmatrix}
1-q & q
\\ q & 1+q
\end{smallmatrix}\right) \otimes \sigma \quad \quad
\forall\sigma\in\B(\H^{B_2}),
\]
where the tensor decomposition $\bbC^4 = \H^{A_2}\otimes\H^{B_2}$
is given above. }
\end{eg}

%\begin{rem}
%{\rm The channel of the previous example was constructed only to
%emphasize that the channel need not be unital for noiseless
%subsystems to exist. This example is mathematically motivated, and
%thus it may seem somewhat artificial from the physical
%perspective. However, the work \cite{CK05} shows that this is the
%typical manner in which noiseless subsystems arise for arbitrary
%channels. In particular, the conditions of Lemma~\ref{NSlemma} can
%be seen to be equivalent to the requirement $\P_\fA \circ \E \circ
%\P_\fA = \E\circ\P_\fA = \E^\prime \otimes id_B$, where
%$\E^\prime$ is some channel on $\H^A$ and $id_B$ is the identity
%channel on $\H^B$. See \cite{CK05} for further discussions on this
%point.}
%\end{rem}

%%%%%%%%%%%%%%%%%%%%%%%%%%%%%%%%%%%%%%%%%%%%%%
%%%%%
\section{Operator Quantum Error Correction}\label{S:unified}
%%%%%%%%%%%%%%%%%%%%%%%%%%%%%%%%%%%%%%%%%%%%%%
%%%%%%%%%%%%%

The unified scheme for quantum error correction consists of a
triple $(\R,\E,\fA)$ where again $\R$ and $\E$ are quantum
operations on some $\B(\H)$, but now $\fA$ is a semigroup in
$\B(\H)$ defined as above with respect to a fixed decomposition
$\H = (\H^A \otimes \H^B) \oplus \K$.

\begin{defn}\label{correctdefn}
{\rm Given such a triple $(\R,\E,\fA)$ we say that the $B$-sector
of $\fA$ is {\it correctable for $\E$} if
\begin{eqnarray}\label{newid}
\big(\Tr_A \circ \P_\fA \circ\R \circ \E \big) (\sigma) =
\Tr_A(\sigma) \qforal \sigma \in \fA.
\end{eqnarray}
}
\end{defn}

In other words, $(\R,\E,\fA)$ is a correctable triple if the
$\H^B$ sector of the semigroup $\fA$ encodes a noiseless subsystem
for the error map $\R\circ\E$. Thus, substituting $\E$ by
$\R\circ\E$ in Lemma~\ref{lemma:generalNS} offers  alternative
equivalent definitions of a correctable triple. Since correctable
codes consist of operator semigroups and algebras, we refer to
this scheme as {\it Operator Quantum Error Correction} (OQEC).
Observe that the standard model for error correction is given by
the particular case in the OQEC model that occurs when $m=\dim\H^A
= 1$. Lemma~\ref{lemma:generalNS} shows that the decoherence-free
subspace  and noiseless subsystem methods are captured in this
model when $\R = {\rm id}$ is the identity channel and,
respectively, $m=1$ and $m\geq 1$.

%These facts are succinctly stated in Table~1. By a ``subspace'' in
%this truth table, we mean the natural identification of a subspace
%$\H^B$ with the operator algebra $\fA \cong \B(\H^B)$ when
%$\dim\H^A = m =1$. Further, the term ``Algebraic NS'' in the table
%is simply meant to refer to the operator algebra subcase of the
%noiseless subsystem notion for arbitrary quantum operations
%discussed at the start of  the previous section.

%\begin{table}

%Table 1: Special Cases of Operator QEC \vspace{0.1in}

%\begin{tabular}{||c||c||} \hline\hline
%$\fA = \mbox{subspace}$ & Standard QEC \\ \hline $\R = id$ &
%Arbitrary NS
%\\ \hline $\R = id \,\, + \,\, \fA = \mbox{algebra}$ & Algebraic
%NS \\ \hline
%$\R = id \,\, + \,\, \fA = \mbox{subspace}$ & DFS \\
%\hline\hline
%\end{tabular}

%\end{table}

While we focus on the general setting of operator semigroups $\fA$
as correctable codes, it is important to note that correctability
of a given $\fA$ is equivalent to the precise correction of the
$\dagger$-algebra
\[
\fA_0 = \one^A \otimes \B(\H^B)
\]
in the following sense. (Note the difference between $\fA_0$ just
defined and $\fA = \{\sigma = \sigma^A \otimes \sigma^B :
\sigma^{A,B} \in \B(H^{A,B})\}$; in the former case the $A$ sector
is restricted to the maximally mixed state while in the latter it
is not.)

\begin{thm}\label{thm:opalgequiv}
Let $\E = \{E_a\}$ be a quantum operation on $\B(\H)$ and let
$\fA$ be a semigroup in $\B(\H)$ as in Eq.~(\ref{eq:semigroup}).
Then the $B$-sector of $\fA$ is correctable for $\E$ if and only
if there is a quantum operation $\R$ on $\B(\H)$ such that
\begin{eqnarray}\label{opalgcorrect}
(\R\circ\E)(\sigma) = \sigma \quad \forall\, \sigma \in \fA_0.
\end{eqnarray}
\end{thm}

\Prf If Eq.~(\ref{opalgcorrect}) holds, then condition~{\it 2.} of
Lemma~\ref{lemma:generalNS} holds for $\R\circ\E$ with $\tau^A =
\one^A$ and hence the $B$-sector of $\fA$ is correctable for $\E$.
For the converse, suppose that condition~{\it 2.} of
Lemma~\ref{lemma:generalNS} holds for $\R\circ\E$. Note that the
map $\Gamma' = \{\frac{1}{\sqrt{m}}P_{kl}\}$ is trace preserving
on $\B(\H^A \otimes \H^B)$. Thus by Lemma~\ref{lemma:map} we have
for all $\sigma^B$,
\begin{eqnarray}\label{opalgequiveqn}
(\Gamma' \circ\R\circ\E) (\one^A \otimes \sigma^B) = \Gamma'
(\tau^A \otimes \sigma^B) \propto \one^A \otimes \sigma^B.
\end{eqnarray}
By trace preservation the proportionality factor must be one, and
hence Eq.~(\ref{opalgcorrect}) is satisfied for $(\Gamma'
\circ\R)\circ\E$. The map $\Gamma^\prime$ may be extended to a
quantum operation on $\B(\H)$ by including the projection
$P_\fA^\perp$ onto $\K$ as a Kraus operator. As this does not
effect the calculation Eq.~(\ref{opalgequiveqn}), the result
follows.
 \bx

We next derive  a testable condition that characterizes
correctable codes for a given channel $\E$ in terms of its error
operators and generalizes Eq.~(\ref{eq:standard}) for the standard
model. We first glean some interesting peripheral information.

\begin{lem}\label{projnfixed}
Let $\E = \{E_a\}$ be a quantum operation on $\B(\H)$ and let $P$
be a projection on $\H$. If $\E(P)=P$, then the range space $\C$
for $P$ is invariant for every $E_a$; that is,
\[
E_a P = P E_a P \quad \forall a.
\]
\end{lem}

\Prf Let $\ket{\psi}$ belong to $\C = P \H$. Then by hypothesis
and the positivity of $\E$, for each $a$ we have
\[
E_a \kb{\psi}{\psi} E_a^\dagger \leq  \sum_b E_b \kb{\psi}{\psi}
E_b^\dagger  = \E(\kb{\psi}{\psi}) \leq \E(P) = P.
\]
Thus $ P^\perp (E_a \kb{\psi}{\psi} E_a^\dagger) P^\perp \leq
P^\perp P P^\perp = 0 $ and so $P^\perp E_a \ket{\psi} =0$. As
both $\ket{\psi}$ and $a$ were arbitrary the result follows.
 \bx

An adjustment of this proof shows that more is true when $\E$ is
contractive ($\E(\one)\leq \one$). Specifically, $\E(P)\leq P$ if
and only if $E_a P = P E_a P$ for all $a$ in this event. In the
special case of unital operations one can further obtain the
following \cite{Kri03a}.

\begin{prop}
 If $\E = \{E_a\}$ is a unital quantum operation
and $P$ is a projector, then $\E(P)=P$ if and only if the range
space for $P$ reduces each $E_a$; that is, $PE_a = E_a P$ for all
$a$.
\end{prop}

We now prove necessary and sufficient conditions for a semigroup
$\fA$ to be correctable for a given error model. Sufficiency was first proven in \cite{NP05}.
We assume that
matrix units $\{P_{kl}\}$ inside $\B(\H^A)\otimes\one^B$ have been
identified as above.

\begin{thm}\label{unifiedthm}
Let $\E = \{E_a\}$ be a quantum operation on $\B(\H)$ and let
$\fA$ be a semigroup in $\B(\H)$ as in Eq.~(\ref{eq:semigroup}).
Then the $B$-sector of $\fA$ is correctable for $\E$ if and only
if for any choice of matrix units $\{P_{kl}\}$ for
$\B(\H^A)\otimes\one^B$ as in Eq.~(\ref{matrixunitdefn}), there
are scalars $\Lambda = ( \lambda_{abkl})$ such that
\begin{eqnarray}\label{condition}
P_{kk} E_a^\dagger E_b P_{ll} = \lambda_{abkl} P_{kl} \quad
\forall a,b,k,l.
\end{eqnarray}
\end{thm}

\Prf To prove necessity, by Theorem~\ref{thm:opalgequiv} we can
assume there is a quantum operation $\R$ on $\B(\H)$ such that
$\R\circ\E$ acts as the identity channel on $\fA_0 = \one^A
\otimes \B(\H^B)\subseteq \B(\H)$. For brevity, we shall first
suppose that $\R = {\rm id}$ is the identity channel.

Let $\C = P_\fA \H$ be the range of the projection $P_\fA = P_{11}
+ \ldots +P_{mm}$. Then since $P_\fA \in \fA_0$ we have $\E(P_\fA)
= P_\fA$ and so Lemma~\ref{projnfixed} gives us $P_\fA E_a |_\C =
E_a |_\C$ for all $a$.

With $\B(\C)$ naturally regarded as imbedded inside $\B(\H)$,
define a completely positive map $\E_\C : \B(\C) \rightarrow
\B(\C)$ via
\[
\sigma \mapsto \E_\C (\sigma) =P_\fA \E(\sigma)|_\C = P_\fA
\E(P_\fA \sigma P_\fA)|_\C
\]
for all $\sigma \in \B(\C)$. Then we have
\begin{eqnarray*}
\sum_a (P_\fA E_a |_\C)^\dagger (P_\fA E_a |_\C) = \sum_a P_\fA
E_a^\dagger E_a|_\C = P_\fA \one_\H |_\C = \one_\C,
\end{eqnarray*}
and so $\E_\C$ defines a quantum operation on $\B(\C)$. Moreover,
$\E_\C$ is unital as $\E_\C (\one_\C) = P_\fA \E (P_\fA)|_\C =
\one_\C$.

Thus by hypothesis and Eq.~(\ref{fixedptthm}) we have
\[
\fA_0|_\C \subseteq \fix(\E_\C)  = \{P_\fA E_a|_\C, P_\fA
E_a^\dagger|_\C\}',
\]
where the latter commutant is computed inside $\B(\C)$. It follows
that
\[
\B(\H^A)\otimes \one^B = ( \fA_0|_\C)' \supseteq  \{ P_\fA E_a
|_\C, P_\fA E_a^\dagger|_\C\}^{\prime\prime} = \ca(\{P_\fA
E_a|_\C\}).
\]
Since the $P_{kl}$ form a set of matrix units that generate
$(P_\fA \fA_0|_\C)^\prime = \B(\H^A)\otimes\one^B$ as a vector
space, there are scalars $\mu_{akl}\in\bbC$ such that
\[
P_{kk} E_a P_{ll} = P_{kk} (P_\fA E_a|_\C) P_{ll} = \mu_{akl}
P_{kl}.
\]

We now turn to the general case and suppose  $\R = \{ R_b \}$. The
noise operators for the operation $\R \circ \E$ are $\{R_b E_a\}$
and thus we may find scalars $\mu_{abkl}$ such that
\[
P_{kk} R_b E_a P_{ll} = \mu_{abkl} P_{kl} \quad \forall a,b,k,l.
\]
Consider the products
\begin{eqnarray*}
\big(P_{kk} R_b E_a P_{ll} \big)^\dagger \big(P_{k'k'} R_b E_{a'}
P_{l'l'}\big) &=& \big( \overline{\mu_{abkl}} P_{lk} \big) \big(
\mu_{a'bk'l'} P_{k'l'} \big) \\
&=& \left\{ \begin{array}{cl} (\overline{\mu_{abkl}}\mu_{a'bkl'}
)P_{ll'} & \mbox{if $k=k'$} \\
0 & \mbox{if $k\neq k'$}
\end{array}\right. .
\end{eqnarray*}
Noting that $\C$ is invariant for the noise operators $R_b E_a$ by
Lemma~\ref{projnfixed}, for fixed $a, a'$ and $l,l'$ we use
$\sum_b R_b^\dagger R_b = \one$ to obtain
\begin{eqnarray*}
\Big( \sum_{b,k} \overline{\mu_{abkl}} \mu_{a'bkl'} \Big) P_{ll'}
&=& \sum_{b,k} \big(P_{ll} E_a^\dagger R_b^\dagger P_{kk} \big)
\big(P_{kk}
R_b E_{a'} P_{l'l'} \big) \\
&=& \sum_b P_{ll} E_a^\dagger R_b^\dagger P_\fA R_b E_{a'}
P_{l'l'} \\
&=&  P_{ll} E_a^\dagger \Big( \sum_b R_b^\dagger R_b \Big) E_{a'} P_{l'l'} \\
&=& P_{ll} E_a^\dagger E_{a'} P_{l'l'}
\end{eqnarray*}
The proof is completed by setting $\lambda_{aa'll'} =\sum_{b,k}
\overline{\mu_{abkl}} \mu_{a'bkl'}$ for all $a,a'$ and $l,l'$.

For sufficiency, let us assume that Eq.~(\ref{condition}) holds.
Let $\sigma_k =\kb{\alpha_k}{\alpha_k}\in\B(\H^A)$, for $1\leq k
\leq m$, and define a quantum operation
$\E_k:\B(\H^B)\rightarrow\B(\H)$ by $ \E_k(\rho^B) \equiv
\E(\sigma_k\otimes\rho^B). $ With $P\equiv P_\fA$ and $E_{a,k}
\equiv E_a P \ket{\alpha_k}$, it follows that $\E_k = \{ E_{a,k}\}
$. We shall find a quantum operation that globally corrects  all
of the errors $E_{a,k}$.

To do this, first note that  we may define a quantum operation
$\E_B:\B(\H^B)\rightarrow\B(\H)$ with error model
\[
\E_B = \big\{ \frac{1}{\sqrt{m}} E_{a,k} : \forall a, \, \forall
1\leq k \leq m\big\}.
\]
Then Eq.~(\ref{condition}) and $P=\sum_k P_{kk}$ give us
\begin{eqnarray*}
\one^B E_{a,k}^\dagger E_{b,l} \one^B &=& \one^B \bra{\alpha_k} P
E_a^\dagger E_b P \ket{\alpha_l} \one^B \\
&=& \sum_{k',l'} \one^B \bra{\alpha_k} P_{k'k'} E_a^\dagger E_b
P_{l'l'} \ket{\alpha_l} \one^B \\
&=& \sum_{k',l'} \lambda_{abk'l'} \, \one^B \bra{\alpha_k}
P_{k'l'} \ket{\alpha_l} \one^B  =  \lambda_{abkl}  \one^B.
\end{eqnarray*}
In particular, Standard QEC implies the existence of a quantum
operation $\R: \B(\H)\rightarrow\B(\H^B)$ such that
$(\R\circ\E_B)(\rho^B) = \rho^B$ for all $\rho^B$.

This implies that
\begin{eqnarray*}
(\R\circ\E)(\one^A\otimes\rho^B) &=& \R \Big( \sum_k \E_k(\rho^B)
\Big) \\ &=& m \, \R \Big( \sum_{k,a} \frac{1}{m} E_{a,k} \rho^B
E_{a,k}^\dagger \Big) \\ &=& m \, \R \circ \E_B (\rho^B) = m
\rho^B.
\end{eqnarray*}
Hence we may define a channel $I_\fA : \B(\H^B)\rightarrow \B(\H)$
via $I_\fA(\rho^B) = \frac{1}{m} (\one^A \otimes \rho^B$). Thus,
on defining $\R' \equiv I_\fA \circ \R$, we obtain
\[
\big( \R' \circ\E\big)(\one^A\otimes\rho^B) = \one^A\otimes\rho^B
\quad \forall\, \rho^B\in\B(\H^B).
\]
The result now follows from an application of
Theorem~\ref{thm:opalgequiv}.
 \bx

\begin{rem}
{\rm The necessity of Eq.~(\ref{condition}) for correction was
initially established in \cite{KLP04}. Here we have provided a new
operator algebra proof based on Eq.~(\ref{fixedptthm}) and
Theorem~\ref{thm:opalgequiv}. In the original draft of this paper,
we established sufficiency of Eq.~(\ref{condition}) up to a set of
technical conditions. More recently, sufficiency was established
in full generality in \cite{NP05}.  In \cite{NP05}, two proofs of
sufficiency were given; the first casts this condition into
information theoretic language, and a sketch was given for the
second. Here we have presented an operator algebra version (based
on Theorem~\ref{thm:opalgequiv}) of the proof of sufficiency
sketched in \cite{NP05}.}
\end{rem}

Let us note that Eq.~(\ref{condition}) is independent of the
choice of basis $\{\ket{\alpha_k}\}$ that define the family
$P_{kl}$ and of the operator-sum representation for $\E$. In
particular, under the changes $\ket{\alpha'_k} = \sum_l
u_{kl}\ket{\alpha_l}$ and $F_a = \sum_b w_{ab} E_b$, the scalars
$\Lambda$ change to $\lambda_{abkl}' = \sum_{a'b'k'l'}
\overline{u}_{kk'}u_{l'l}\overline{w}_{aa'}w_{bb'}
\lambda_{abkl}$.

Equation~(\ref{condition}) generalizes the quantum error
correction  condition Eq.~(\ref{eq:standard}) to the case where
information is encoded in operators, not necessarily restricted to
act on a fixed code subspace $\C$. However, observe that setting
$k= l$ in Eq.~(\ref{condition}) gives the standard error
correction condition Eq.~(\ref{eq:standard}) with $P_\C = P_{kk}$.
This leads to the following result.

\begin{thm}
If $(\R,\E,\fA)$ is a correctable triple for some semigroup $\fA$
defined as in Eq.~(\ref{eq:semigroup}), then
$(\P_k\circ\R,\E,P_{kk}\fA P_{kk})$ is a correctable triple
according to the standard definition Eq.~(\ref{eq:standard}),
where $P_{kk}$ is any minimal reducing projection of $\fA_0
=\one^A \otimes \B(\H^B)$, and the map $\P_k$ is defined by
$\P_k(\cdot) = \sum_l P_{kl}(\cdot )P_{kl}^\dagger$.
\label{thm:standard}
\end{thm}

\Prf Let $\sigma\in\kb{\alpha_k}{\alpha_k}\otimes\B(\H^B)$, so
that $\sigma = P_{kk}\sigma P_{kk}$. Let $\E=\{E_a\}$ and $\R =
\{R_b\}$. By Theorem~\ref{thm:NS} there are scalars
$\lambda_{abkl}$ such that $P_{kk} R_b E_a P_{ll} = \lambda_{abkl}
P_{kl}$ $\forall\, a,b,k,l$. It follows that
\begin{eqnarray*}
(\P_k\circ\R\circ\E)(\sigma) &=& \sum_{a,b,l} P_{kl} R_b E_a
P_{kk} \sigma P_{kk}
E_a^\dagger R_b^\dagger P_{lk} \\
&=& \sum_{a,b,l} (\lambda_{ablk}P_{kk}) \sigma
(\overline{\lambda_{ablk}}P_{kk})
\\ &=& \Big( \sum_{a,b,l} |\lambda_{ablk}|^2\Big) \sigma .
\end{eqnarray*}
Thus $(\P_k\circ\R\circ\E)(\sigma)\propto \sigma$ for all
$\sigma\in \kb{\alpha_k}{\alpha_k}\otimes\B(\H^B)$, the
proportionality factor independent of $\sigma$. In fact, this
factor is one. To see this, fix $k$ and note that
Theorem~\ref{thm:NS} shows that
\[
R_bE_aP_{kk} = R_bE_a P_\fA P_{kk} = P_\fA R_b E_a P_\fA P_{kk} =
P_\fA R_b E_a P_{kk} \quad \forall\, a,b.
\]
Hence, trace preservation of $\R\circ\E$ yields
\begin{eqnarray*}
\big(\sum_{a,b,l} |\lambda_{ablk}|^2\big) P_{kk} &=& \sum_{a,b,l}
(P_{kk} E_a^\dagger R_b^\dagger P_{ll})( P_{ll} R_b E_a P_{kk}) \\
&=& P_{kk} \big(
\sum_{a,b} E_a^\dagger R_b^\dagger P_\fA R_b E_a \big) P_{kk} \\
&=& P_{kk}\big( \sum_{a,b} E_a^\dagger R_b^\dagger R_b E_a \big)
P_{kk} = P_{kk}.
\end{eqnarray*}
As $k$ was arbitrary, the result follows.
 \bx

\begin{rem}
{\rm Theorem~\ref{thm:standard} has important consequences. Given
a map $\E$, the existence of a correctable code subspace $\C$ ---
captured by the standard error correction condition
Eq.~(\ref{eq:standard})
--- is a prerequisite to the existence of any known type of error
correction or prevention scheme (including the generalizations
introduced here and in \cite{KLP04}). Moreover,
Theorem~\ref{thm:standard} shows how to transform any one of these
error correction or prevention techniques into a standard error
correction scheme. However, while OQEC does not lead to new
families of codes, it does allow for simpler correction
procedures. See \cite{Bac05,Pou05} for further discussions on this
point.}
\end{rem}

\begin{rem}
{\rm As a special case, Theorem~\ref{thm:standard}  demonstrates
that to every noiseless subsystem, there is an associated QEC code
obtained by projecting the $A$-sector to a pure state. This is
complementary to Theorem~6 of \cite{KLV00a} which demonstrates
that every QEC scheme composed of a triple $(\R,\E,\C)$ arises as
a noiseless subsystem of the map $\E \circ \R$.}
\end{rem}

%%%%%%%%%%%%%%%%%%%%%%%%%%%%%%%%%%%%%%%%%%%%%%%%%%%%%%%%%
%\section{Ampliation Channels}\label{S:ampliations}
%%%%%%%%%%%%%%%%%%%%%%%%%%%%%%%%%%%%%%%%%%%%%%%%%%%%%%%%%

We conclude this section by exhibiting the 2-qubit case of a new
class of quantum channels, together with correctable subsystems,
that is covered by OQEC, but for which the recovery operation does
not fit into the Standard QEC protocol.

First, let us recall briefly that the motivating class of channels
$\E = \{E_a\}$ which satisfy Eq.~(\ref{eq:standard}) occur when
the restrictions $E_a|_{P_\C\H}= E_a|_\C$ of the error operators
to $\C$ are scalar multiples of unitary operators $U_a$ such that
the subspaces $U_a\C$ are mutually orthogonal. In fact, this case
describes any error model that satisfies Eq.~(\ref{eq:standard}),
up to a linear transformation of the error operators. In this
situation the positive scalar matrix $\Lambda$ is diagonal. A
correction operation here may be constructed by an application of
the measurement operation determined by the subspaces $U_a\C$,
followed by the reversals of the corresponding restricted
unitaries $U_a P_\C$. Specifically, if $P_a$ is the projection of
$\H$ onto $U_a\C$, then $\R = \{U_a^\dagger P_a\}$ satisfies
Eq.~(\ref{reverse}) for $\E$ on $\C$. The following is a
generalization of this class of channels to the OQEC setting. For
clarity we focus on the 2-qubit case.

\begin{eg}\label{lasteg}
{\rm Let $\{\ket{a},\ket{b},\ket{a'},\ket{b'}\}$ and
$\{\ket{a_1},\ket{b_1},\ket{a_2},\ket{b_2}\}$ be two orthonormal
bases for $\bbC^4$. Let $P_1$ be the projection onto $\spn\{
\ket{a},\ket{b}\}$ and $P_2$ the projection onto $\spn\{
\ket{a'},\ket{b'}\}$. Let $Q_i$, $i=1,2$, be the projection onto
$\spn\{\ket{a_i},\ket{b_i}\}$. Define operators $U_1$,
$U_1^\prime$, $U_2$, $U_2^\prime$  on $\bbC^4$ as follows:
\[
\left\{ \begin{array}{rcl}
U_1 \ket{a} &=& \ket{a_1}  \\
U_1 \ket{b} &=& \ket{b_1}  \\
U_1' \ket{a'} &=& \ket{a_1}  \\
U_1' \ket{b'} &=& \ket{b_1}
\end{array}\right.
\quad\quad \left\{ \begin{array}{rcl}
U_2 \ket{a} &=& \ket{a_2}  \\
U_2 \ket{b} &=& \ket{b_2}  \\
U_2' \ket{a'} &=& \ket{a_2}  \\
U_2' \ket{b'} &=& \ket{b_2}
\end{array}\right.,
\]
and put $U_1P_2 \equiv U_1^\prime P_1 \equiv U_2P_2 \equiv
U_2^\prime P_1 \equiv 0$. Then these operators are ``partial
isometries'' and satisfy $U_1=U_1P_1$, $U_1'=U_1'P_2$,
$U_2=U_2P_1$, $U_2'=U_2' P_2$. The operators $\E = \{E_1,E_2\}$
define a quantum channel where
\[
E_1 = \frac{1}{\sqrt{2}} \big( U_1P_1 + U_1'P_2 \big)
\]
\[
E_2 = \frac{1}{\sqrt{2}} \big( U_2P_1 - U_2'P_2 \big).
\]
The action of $E_1$ and $E_2$ is indicated in Figure~1.

\begin{figure}[h]\caption{}

\setlength{\unitlength}{.007in}

\begin{picture}(100,340)(150,0)

\put(20,80){\oval(60,120)}

\put(60,80){\line(1,0){20}}

\put(90,80){\line(1,0){20}}

\put(120,80){\line(1,0){20}}

\put(150,80){\line(1,0){20}}

\put(180,80){\line(1,0){20}}

\put(210,80){\line(1,0){20}}

\put(240,80){\line(1,0){20}}

\put(270,80){\line(1,0){20}}

\put(300,80){\line(1,0){20}}

\put(330,80){\vector(1,0){10}}

\put(60,80){\vector(2,1){280}}

\put(100,125){$E_1$}

\put(100,185){$E_2$}

\put(10,50){$\ket{b'}$}

\put(10,100){$\ket{a'}$}

\put(-60,70){$P_2$}

\put(-60,230){$P_1$}

\put(20,240){\oval(60,120)}

\put(60,240){\vector(1,0){280}}

\put(60,240){\line(2,-1){20}}

\put(90,225){\line(2,-1){20}}

\put(120,210){\line(2,-1){20}}

\put(150,195){\line(2,-1){20}}

\put(180,180){\line(2,-1){20}}

\put(210,165){\line(2,-1){20}}

\put(240,150){\line(2,-1){20}}

\put(270,135){\line(2,-1){20}}

\put(300,120){\vector(2,-1){30}}

\put(10,210){$\ket{b}$}

\put(10,260){$\ket{a}$}

\put(380,240){\oval(60,120)}

\put(380,80){\oval(60,120)}

\put(430,70){$Q_2$}

\put(430,230){$Q_1$}

\put(367,50){$\ket{b_2}$}

\put(367,100){$\ket{a_2}$}

\put(367,210){$\ket{b_1}$}

\put(367,260){$\ket{a_1}$}

\put(190,260){$E_1$}

\put(190,45){$E_2$}

\end{picture}

\end{figure}

Here the matrix units are given by
\[
P_{1} =P_{11} =\ket{a}\bra{a} + \ket{b}\bra{b}
\]
\[
P_2  = P_{22} =\ket{a'}\bra{a'} + \ket{b'}\bra{b'}
\]
\[
P_{12} =\ket{a}\bra{a'} + \ket{b}\bra{b'}
\]
\[
P_{21} =\ket{a'}\bra{a} + \ket{b'}\bra{b}.
\]
For trace preservation, observe that
\begin{eqnarray*}
E_1^\dagger E_1 &=& \frac{1}{2} \big( P_1 U_1^\dagger + P_2
(U_1')^\dagger \big) \big( U_1 P_1 + U_1'P_2 \big) \\ &=&
\frac{1}{2} \big( P_{11} + P_{12} + P_{21} + P_{22} \big).
\end{eqnarray*}
Similarly, we compute
\[
E_2^\dagger E_2 = \frac{1}{2} \big( P_{11} - P_{12} - P_{21} +
P_{22} \big).
\]
Thus we have $E_1^\dagger E_1 + E_2^\dagger E_2 = P_{11} + P_{22}
= \one_4$. Equations~(\ref{condition}) are computed as follows:
\[
P_k E_i^\dagger E_i P_k = \frac{1}{2} P_k \qfor i,k = 1,2,
\]
\[
P_k E_i^\dagger E_j P_l = 0 \qfor i\neq j \qand k,l = 1,2,
\]
\[
P_1 E_1^\dagger E_1 P_2 = \frac{1}{2} P_{12} = \big( \frac{1}{2}
P_{21}\big)^\dagger = \big( P_2 E_1^\dagger E_1 P_1 \big)^\dagger,
\]
\[
P_1 E_2^\dagger E_2 P_2 = \frac{-1}{2} P_{12} = \big( \frac{-1}{2}
P_{21}\big)^\dagger = \big( P_2 E_2^\dagger E_2 P_1 \big)^\dagger.
\]

Define
\[
V_{11}=U_1P_1, \quad V_{12}=U_1'P_2,\quad V_{21}=U_2P_1,\quad
V_{22}=U_2'P_2
\]
and observe that
\[
V_{11} V_{11}^\dagger = U_1P_1U_1^\dagger = Q_1 = U_1^\prime
P_2(U_1^\prime)^\dagger = V_{12}V_{12}^\dagger
\]
\[
V_{21}V_{21}^\dagger = U_2 P_1 U_2^\dagger = Q_2 = U_2' P_2
(U_2')^\dagger = V_{22} V_{22}^\dagger.
\]
Then a calculation shows that the channel
\[
\R = \Big\{ \frac{1}{\sqrt{2}} V_{jk}^\dagger Q_j : 1\leq j,k \leq
2 \Big\}
\]
corrects for all errors induced by $\E$ on $\fA_0 \cong
\one_2\otimes \M_2$. Specifically, $(\R\circ\E)(\sigma) = \sigma$
for all $\sigma\in\B(\bbC^4)$ which have a matrix
representation of the form $\sigma = \begin{spmatrix} \sigma_1 & 0 \\
0 & \sigma_1 \end{spmatrix}$, $\sigma_1\in \M_2$, with respect to
the ordered basis $\{\ket{a},\ket{b},\ket{a'},\ket{b'}\}$ for
$\bbC^4$. That is, $(\R\circ\E)(\sigma)=\sigma$ for all
$\alpha_{11},\alpha_{12},\alpha_{21},\alpha_{22}\in\bbC$ and all
\begin{eqnarray*}
\sigma &=& \alpha_{11}\big(\kb{a}{a} + \kb{a'}{a'}\big) +
\alpha_{12}\big(\kb{a}{b} + \kb{a'}{b'}\big) \\ & &
+\alpha_{21}\big(\kb{b}{a} + \kb{b'}{a'}\big)
+\alpha_{22}\big(\kb{b}{b} + \kb{b'}{b'}\big).
\end{eqnarray*}
Thus $\R$ corrects all $\sigma= \one_2\otimes\sigma_1$ that  are
``equally balanced'' with respect to the standard bases for the
ranges of $P_1$ and $P_2$. Further, by
Theorem~\ref{thm:opalgequiv} we know $\R$ corrects the associated
semigroup $\fA$ in the sense of Definition~\ref{correctdefn}. }
\end{eg}

\begin{rem}
{\rm We note that recent work \cite{Bac05} presents physically
motivated examples in which correction of subsystems is
accomplished within the OQEC framework. Furthermore, a general
class of recovery procedures based on the stabilizer formalism was
recently presented in \cite{Pou05}. In particular, this work
builds on OQEC to demonstrate how certain stabilizer codes can be
simplified by incorporating gauge qubits. These have the effect of
reducing the number of syndrome measurements required to correct
the error map and extend the class of physical realizations  of
the logical operations on the encoded data. }
\end{rem}

%%%%%%%%%%%%%%%%%%%%%%%%%%%%%%%%%%%%%%%%%%%%%%%%%%%%%%%%%
\section{Unitarily Noiseless Subsystems}\label{S:fixedunitary}
%%%%%%%%%%%%%%%%%%%%%%%%%%%%%%%%%%%%%%%%%%%%%%%%%%%%%%%%%

In this section we discuss error triples $(\R,\E,\fA)$ such that
the restriction of $\R$ to $\E(\fA)$ is a unitary operation.
Consideration of this case leads to a generalization  of the
noiseless subsystem protocol that falls under the OQEC umbrella.
Let us first consider a direct generalization of the fixed point
set algebraic approach as in Eq.~(\ref{fixedptthm}). Here we
have the equation
\begin{eqnarray}\label{unitaryevolve}
\E (\sigma) = U \sigma U^\dagger \quad \forall\,\sigma\in\fA_0 =
\one^A \otimes \B(\H^B),
\end{eqnarray}
for some unitary operator $U$.  When $\fA_0$ satisfies
Eq.~(\ref{unitaryevolve}) for a unitary $U$ we shall say that
$\fA_0$ is a {\it unitarily noiseless subsystem} (UNS) for $\E$.
Of course, a subsystem $\fA_0$ that satisfies
Eq.~(\ref{unitaryevolve}) is not noiseless, but it may be easily
corrected by applying the reversal operation $U^\dagger(\cdot)U$.
As we indicate below, this can lead to new non-trivial correctable
subsystems not obtained under the noiseless subsystem regime. If
$\E$ is a unital operation, it is possible to explicitly compute
all UNS's for $\E$.

\begin{thm}\label{unitarythm}
If $\E = \{E_a\}$ is a unital quantum operation on $\B(\H)$ and
$U$ is a unitary on $\H$, then the corresponding unitarily
noiseless subsystem $\fA_0$ is equal to the commutant of the
operators $\{ U^\dagger E_a \}$;
\begin{eqnarray*}\label{unitarycommutant}
\fA_0 &=& \big\{ \sigma\in\B(\H): \E (\sigma)=\sum_a E_a\sigma
E_a^\dagger = U \sigma U^\dagger \big\} \\ &=& \big\{ U^\dagger
E_a \big\}^\prime.
\end{eqnarray*}
\end{thm}

\Prf The set of $\sigma$ that satisfy Eq.~(\ref{unitaryevolve}) is
equal to the set of $\sigma$ that satisfy $U^\dagger \E (\sigma) U
= \sigma$. Thus, here we are considering the fixed point set for
the unital operation $U^\dagger \E (\cdot) U$, which has noise
operators $\{U^\dagger E_a\}$. The result now follows from
Eq.~(\ref{fixedptthm}).
 \bx

Let us consider a simple example of how this scheme can be used to
identify new correctable codes for a given channel.

\begin{eg}
{\rm Let $Z_1 = Z\otimes \one_2$ and $Z_2 = \one_2\otimes Z$ with
the Pauli matrix $Z = \begin{spmatrix} 1 & 0 \\ 0 & -1
\end{spmatrix}$. Then, with respect to the standard orthonormal basis
$\{\ket{00}, \ket{01}, \ket{10}, \ket{11}\}$ for $\bbC^4$, we have
\begin{eqnarray*}
\{Z_1,Z_2\}' &=& \left\{ \left( \begin{matrix} a & 0 & 0 & 0 \\
 0 & b & 0 & 0 \\ 0 & 0 & c & 0 \\ 0 & 0 & 0 & d \end{matrix}
 \right) : a,b,c,d\in \bbC \right\},
 \end{eqnarray*}
Hence there are no non-trivial noiseless subsystems for the
corresponding channel $\E = \{ Z_1,Z_2\}$. However, if we let
$U\in\B(\bbC^4)$ be the unitary
\[
U \ket{ij} = \left\{ \begin{array}{cl} \ket{ij} & \mbox{if $i\neq
1$ or $j\neq 1$} \\ -\ket{11} & \mbox{if $i=1$ and $j=1$}
\end{array}\right.,
\]
then we compute
\begin{eqnarray*}
\{U^\dagger Z_1,U^\dagger Z_2\}' = \left\{ \left( \begin{matrix} a & 0 & 0 & b \\
 0 & c & 0 & 0 \\ 0 & 0 & d & 0 \\ e & 0 & 0 & f \end{matrix}
 \right) : a,b,c,d,e,f\in\bbC \right\}.
 \end{eqnarray*}
In particular, the $\dagger$-algebra $\fA_0 = \{U^\dagger Z_i\}'$
is unitarily equivalent to $\fA_0 \cong \M_2 \oplus \bbC \oplus
\bbC$. Thus, a single qubit code subspace may be corrected.
Specifically, all operators $\sigma\in\fA_0$ may be corrected by
applying $U^\dagger(\cdot) U$ since they satisfy $\E(\sigma) = U
\sigma U^\dagger$.}
\end{eg}

In a similar manner we can extend this discussion to the case of
noiseless subsystems for arbitrary quantum operations. The
analogue of Eq.~(\ref{unitaryevolve}) in this case is
\begin{eqnarray}\label{guns}
\forall\sigma^A\ \forall\sigma^B,\ \exists \tau^A\ :\
\E(\sigma^A\otimes\sigma^B) = U (\tau^A\otimes \sigma^B
)U^\dagger,
\end{eqnarray}
where $U$ is a fixed unitary on $\H$. In effect, this is the
special case of the OQEC formulation Eq.~(\ref{newid}) where the
recovery $\R$ is unitary.  In this context the conditions of
Lemma~\ref{lemma:generalNS} yield the following.

\begin{thm}\label{gunsthm}
Given a fixed decomposition $\H = (\H^A\otimes\H^B) \oplus\K$, a
map $\E$ on $\B(\H)$ and a unitary $U$ on $\H$, the following
three conditions are equivalent:
\begin{enumerate}
\item Eq.~(\ref{guns}) is satisfied. \item $ \forall\sigma^B,\
\exists \tau^A\ :\ \E(\one^A\otimes\sigma^B) = U( \tau^A \otimes
\sigma^B) U^\dagger$ . \item $\forall\sigma\in \fA\ :\
\big(\Tr_A \circ\P_\fA\circ \U^{-1}\circ\E\big)(\sigma)
=\Tr_A(\sigma)$.
\end{enumerate}
where $\U^{-1}(\cdot) = U^\dagger (\cdot) U$.
\end{thm}

\section{Conclusion}\label{S:conclusion}
%%%%%%%%%%%%%%%%%%%%%%%%%%%%%%%%%%%%%%%%%%%%%%%%%%%%%%%%%

We have presented a detailed analysis of the OQEC formalism for
error correction in quantum computing. This approach provides a
unified framework for investigations into both active and passive
error correction techniques. Fundamentally, we have generalized
the setting for correction from states to operators. The condition
from standard quantum error correction was shown to be necessary
for any of these schemes to be feasible. Included in this
formalism is a scheme for identifying noiseless subsystems that
applies to arbitrary (not necessarily unital) quantum operations.
We also introduced the notion of unitarily noiseless subsystems as
a natural relaxation of the noiseless subsystem condition.
%In the updated draft of this
%paper, we have shown that this coincides with the central notion
%in the formulation of the quantum computer condition from
%\cite{GHT05}.

%%%%%%%%%%%%%%%%%%%%%%%%%%%%%%%%%%%%%%%%%%%%%%
%%%%%%%%%%%%%
%%%%%%%%%%%%%%%%%%%%%%%%%%%%%%%%%%%%%%%%%%%%%%
%%%%%%%%%%%%%

\vspace{0.1in}

{\noindent}{\it Acknowledgements.} We thank  Man-Duen Choi,
Michael Nielsen, Harold Ollivier, Rob Spekkens and our other
colleagues for helpful discussions. This work was supported in
part by funding from NSERC, CIAR, MITACS, NATEQ, and ARDA.

%%%%%%%%%%%%%%%%%%%%%%% REFERENCES
%%%%%%%%%%%%%%%%%%%%%%%%%%%%


\begin{thebibliography}{30}
\expandafter\ifx\csname
natexlab\endcsname\relax\def\natexlab#1{#1}\fi
\expandafter\ifx\csname bibnamefont\endcsname\relax
  \def\bibnamefont#1{#1}\fi
\expandafter\ifx\csname bibfnamefont\endcsname\relax
  \def\bibfnamefont#1{#1}\fi
\expandafter\ifx\csname citenamefont\endcsname\relax
  \def\citenamefont#1{#1}\fi
\expandafter\ifx\csname url\endcsname\relax
  \def\url#1{\texttt{#1}}\fi
\expandafter\ifx\csname
urlprefix\endcsname\relax\def\urlprefix{URL }\fi
\providecommand{\bibinfo}[2]{#2}
\providecommand{\eprint}[2][]{\url{#2}}


\bibitem{KLP04}
\bibinfo{author}{\bibfnamefont{D.~W.}~\bibnamefont{Kribs}},
  \bibinfo{author}{\bibfnamefont{R.}~\bibnamefont{Laflamme}}, \bibnamefont{and}
  \bibinfo{author}{\bibfnamefont{D.}~\bibnamefont{Poulin}},
   \bibinfo{journal}{Phys. Rev. Lett.} \textbf{\bibinfo{volume}{94}},
  \bibinfo{pages}{180501} (\bibinfo{year}{2005}).

\bibitem{Sho95a}

\bibinfo{author}{\bibfnamefont{P.~W.} \bibnamefont{Shor}},
  \bibinfo{journal}{Phys. Rev. A} \textbf{\bibinfo{volume}{52}},
  \bibinfo{pages}{R2493} (\bibinfo{year}{1995}).

\bibitem{Ste96a}
\bibinfo{author}{\bibfnamefont{A.~M.} \bibnamefont{Steane}},
  \bibinfo{journal}{Phys. Rev. Lett.} \textbf{\bibinfo{volume}{77}},
  \bibinfo{pages}{793} (\bibinfo{year}{1996}).

  \bibitem{BDSW96a}
\bibinfo{author}{\bibfnamefont{C.~H.} \bibnamefont{Bennett}},
  \bibinfo{author}{\bibfnamefont{D.~P.} \bibnamefont{DiVincenzo}},
  \bibinfo{author}{\bibfnamefont{J.~A.} \bibnamefont{Smolin}},
  \bibnamefont{and} \bibinfo{author}{\bibfnamefont{W.~K.}
  \bibnamefont{Wootters}}, \bibinfo{journal}{Phys. Rev. A}
  \textbf{\bibinfo{volume}{54}}, \bibinfo{pages}{3824} (\bibinfo{year}{1996}).

\bibitem{KL97a}
\bibinfo{author}{\bibfnamefont{E.}~\bibnamefont{Knill}} \bibnamefont{and}
  \bibinfo{author}{\bibfnamefont{R.}~\bibnamefont{Laflamme}},
  \bibinfo{journal}{Phys. Rev. {A}} \textbf{\bibinfo{volume}{55}},
  \bibinfo{pages}{900} (\bibinfo{year}{1997}).

\bibitem{PSE96}
\bibinfo{author}{\bibfnamefont{G.M.}~\bibnamefont{Palma}},
\bibinfo{author}{\bibfnamefont{K.-A.}~\bibnamefont{Suominen}} \bibnamefont{and}
  \bibinfo{author}{\bibfnamefont{A.}~\bibnamefont{Ekert}},
  \bibinfo{journal}{Proc. Royal Soc. A} \textbf{\bibinfo{volume}{452}},
  \bibinfo{pages}{567} (\bibinfo{year}{1996}).

\bibitem{DG97c}
\bibinfo{author}{\bibfnamefont{L.-M.} \bibnamefont{Duan}} \bibnamefont{and}
  \bibinfo{author}{\bibfnamefont{G.-C.} \bibnamefont{Guo}},
  \bibinfo{journal}{Phys. Rev. Lett.} \textbf{\bibinfo{volume}{79}},
  \bibinfo{pages}{1953} (\bibinfo{year}{1997}).

\bibitem{ZR97c}
\bibinfo{author}{\bibfnamefont{P.}~\bibnamefont{Zanardi}} \bibnamefont{and}
  \bibinfo{author}{\bibfnamefont{M.}~\bibnamefont{Rasetti}},
  \bibinfo{journal}{Phys. Rev. Lett.} \textbf{\bibinfo{volume}{79}},
  \bibinfo{pages}{3306} (\bibinfo{year}{1997}).

\bibitem{LCW98a}
\bibinfo{author}{\bibfnamefont{D.A.}~\bibnamefont{Lidar}},
  \bibinfo{author}{\bibfnamefont{I.L.}~\bibnamefont{Chuang}}, \bibnamefont{and}
  \bibinfo{author}{\bibfnamefont{K.B.}~\bibnamefont{Whaley}},
  \bibinfo{journal}{Phys. Rev. Lett.} \textbf{\bibinfo{volume}{81}},
  \bibinfo{pages}{2594} (\bibinfo{year}{1998}).

\bibitem{KLV00a}
\bibinfo{author}{\bibfnamefont{E.}~\bibnamefont{Knill}},
  \bibinfo{author}{\bibfnamefont{R.}~\bibnamefont{Laflamme}}, \bibnamefont{and}
  \bibinfo{author}{\bibfnamefont{L.}~\bibnamefont{Viola}},
  \bibinfo{journal}{Phys. Rev. Lett.} \textbf{\bibinfo{volume}{84}},
  \bibinfo{pages}{2525} (\bibinfo{year}{2000}).

\bibitem{Zan01b}
\bibinfo{author}{\bibfnamefont{P.}~\bibnamefont{Zanardi}},
  \bibinfo{journal}{Phys. Rev. A} \textbf{\bibinfo{volume}{63}},
  \bibinfo{pages}{12301} (\bibinfo{year}{2001}).

\bibitem{KBLW01a}
\bibinfo{author}{\bibfnamefont{J.}~\bibnamefont{Kempe}},
  \bibinfo{author}{\bibfnamefont{D.}~\bibnamefont{Bacon}},
  \bibinfo{author}{\bibfnamefont{D.~A.} \bibnamefont{Lidar}}, \bibnamefont{and}
  \bibinfo{author}{\bibfnamefont{K.~B.} \bibnamefont{Whaley}},
  \bibinfo{journal}{Phys. Rev. A} \textbf{\bibinfo{volume}{63}},
  \bibinfo{pages}{42307} (\bibinfo{year}{2001}).

%\bibitem{GHT05}
%\bibinfo{author}{\bibfnamefont{G.}~\bibnamefont{Gilbert}},
%\bibinfo{author}{\bibfnamefont{M.}~\bibnamefont{Hamrick}},
% \bibnamefont{and}
%  \bibinfo{author}{\bibfnamefont{F.~J.}~\bibnamefont{Thayer}},
%  \bibinfo{journal}{arxiv.org/quant-ph/0507141}.

\bibitem{Arv76a}
\bibinfo{author}{\bibfnamefont{W.}~\bibnamefont{Arveson}},
  \bibinfo{publisher}{Springer - Verlag, New York - Heidelberg},
  \bibinfo{year}{1976}.

\bibitem{Dav96a}
\bibinfo{author}{\bibfnamefont{K.~R.}~\bibnamefont{Davidson}},
  \bibinfo{publisher}{Amer. Math. Soc., Providence},
  \bibinfo{year}{1996}.

  \bibitem{Tak79a}
\bibinfo{author}{\bibfnamefont{M.}~\bibnamefont{Takesaki}},
  \bibinfo{publisher}{Springer - Verlag, New York - Heidelberg},
  \bibinfo{year}{1979}.

\bibitem{Kri03a}
\bibinfo{author}{\bibfnamefont{D.~W.} \bibnamefont{Kribs}},
  \bibinfo{journal}{Proc. Edin. Math. Soc.} \textbf{\bibinfo{volume}{46}}
  (\bibinfo{year}{2003}).


\bibitem{CK05}
\bibinfo{author}{\bibfnamefont{M.~D.}~\bibnamefont{Choi}} \bibnamefont{and}
  \bibinfo{author}{\bibfnamefont{D.~W.}~\bibnamefont{Kribs}},
  \bibinfo{journal}{Phys. Rev. Lett., to appear}.


\bibitem{NP05}
\bibinfo{author}{\bibfnamefont{M.~A.}~\bibnamefont{Nielsen}} \bibnamefont{and}
  \bibinfo{author}{\bibfnamefont{D.}~\bibnamefont{Poulin}},
  \bibinfo{journal}{arxiv.org/quant-ph/0506069}.

\bibitem{Bac05}
\bibinfo{author}{\bibfnamefont{D.}~\bibnamefont{Bacon}},
  \bibinfo{journal}{arxiv.org/quant-ph/0506023}.

\bibitem{Pou05}
\bibinfo{author}{\bibfnamefont{D.}~\bibnamefont{Poulin}},
  \bibinfo{journal}{Phys. Rev. Lett.} \textbf{\bibinfo{volume}{95}},
  \bibinfo{pages}{230504} (\bibinfo{year}{2005}).

\end{thebibliography}
\end{document}